\documentclass[prl,twocolumn,preprintnumbers,amsmath,amssymb,floatfixX]{revtex4-1}

\usepackage{dcolumn}
\usepackage{graphicx}
\usepackage{epstopdf}
\usepackage{natbib}

\begin{document}
\title{Wigner Crystals of $^{229}$Th for Optical Excitation of the Nuclear Isomer}
\author{C.\,J. Campbell}
\affiliation{
School of Physics, Georgia Institute of Technology, Atlanta, Georgia 30332-0430, USA}

\author{A.\,G. Radnaev}
\affiliation {
School of Physics, Georgia Institute of Technology, Atlanta, Georgia 30332-0430, USA}

\author{A. Kuzmich}
\email{alex.kuzmich@physics.gatech.edu}
\affiliation {
School of Physics, Georgia Institute of Technology, Atlanta, Georgia 30332-0430, USA}

\date{June 2, 2011}
\begin{abstract}
We have produced laser-cooled Wigner crystals of $^{229}$Th$^{3+}$ in a linear Paul trap.  The magnetic dipole ($A$) and electric quadrupole ($B$) hyperfine constants for four low-lying electronic levels and the relative isotope shifts with respect to $^{232}$Th$^{3+}$ for three low-lying optical transitions are measured.  Using the hyperfine $B$ constants in conjunction with prior atomic structure calculations, a new value of the spectroscopic nuclear electric quadrupole moment $Q$ = 3.11(16) eb is deduced.  These results are a step towards optical excitation of the low-lying isomer level in the $^{229}$Th nucleus.
\end{abstract}

\maketitle

Atomic nuclei usually possess excitation energies in the keV to MeV range. A notable exception is the $^{229}$Th nucleus where the energy splitting of the ground state doublet is less than 10 eV \cite{Kroger76,Reich90,Helmer94,Beck07}. This fortuitous coincidence offers prospects for a better clock and for a test of temporal variation of fundamental constants at an unprecedented level of precision \cite{Peik03,Flambaum06}.  Here we report laser cooling and crystallization of $^{229}$Th$^{3+}$ in a linear Paul trap. The monovalent character of the Th$^{3+}$ ion is favorable for the isomer level search based on an electron-bridge process \cite{Campbell09}. Once the isomer energy value is measured, either in our system or using one of other approaches \cite{Porsev10crystal,Rellergert10}, a single trapped, cold $^{229}$Th$^{3+}$ ion is expected to be an ideal system to take advantage of this remarkable nucleus.

\begin{figure}[b]
\includegraphics[scale=0.87]{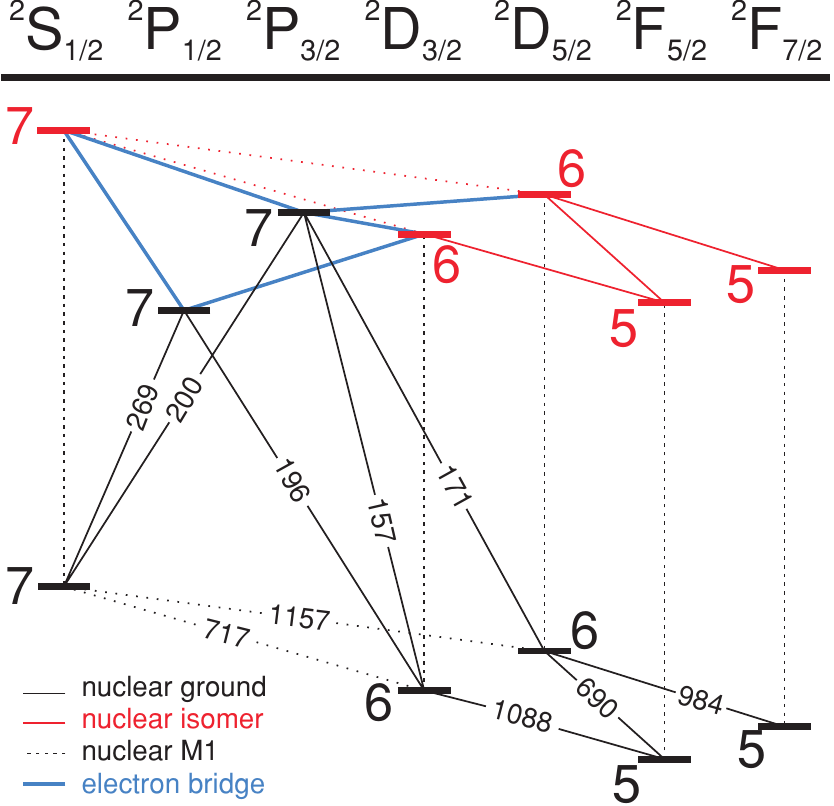}
\caption{\label{fig:grotrian} A diagram of electronic energy levels and electric-dipole transitions including both nuclear ground and excited isomeric manifolds. Direct nuclear magnetic dipole ($M1$) decay channels and electron-bridge pathways suitable for the isomer search are shown.  Optical transition wavelengths are  in nm and the integers near atomic levels indicate principle quantum numbers.}
\end{figure}

Because to the intrinsic narrowness of the nuclear transition and the superb isolation of the nucleus from external fields, $^{229}$Th may form the basis of a next-generation optical clock  \cite{Peik03}. To this end, a convenient set of electronic transitions in the triply charged form of the Th atom may be utilized for optical preparation and readout of the nuclear ground and isomer manifolds. The clock levels may be associated with the metastable $7S$ electronic levels of the ground and isomer manifolds, while allowed electric dipole transitions from the $F$ states are used for laser cooling of the ion and for optical pumping preparation of the clock, Fig. \ref{fig:grotrian}.

Importantly, the ground-isomer nuclear transition may be used to search for temporal variation of fundamental constants by monitoring the ratio of the frequencies of the nuclear transition and that of an atomic (electronic) clock transition, e.g. in Al$^+$ or Sr.  Flambaum {\it et al}. have predicted that such a frequency ratio is likely to be supremely sensitive to relative variation of the strong interaction parameter and the fine structure constant, with enhancement factor $K$ possibly reaching as high as $10^6$ \cite{Flambaum06,He08,Flambaum09}.  Although the complexity of this nuclear system makes a theoretical evaluation of $K$ difficult \cite{Hayes08,Litvinova09}, optical and microwave spectroscopy of the ground and isomer level manifolds should provide a reliable empirical determination of the enhancement factor with little or no dependence on the nuclear model assumed \cite{Berengut09}.

\begin{figure}[t]
\includegraphics[scale=.9]{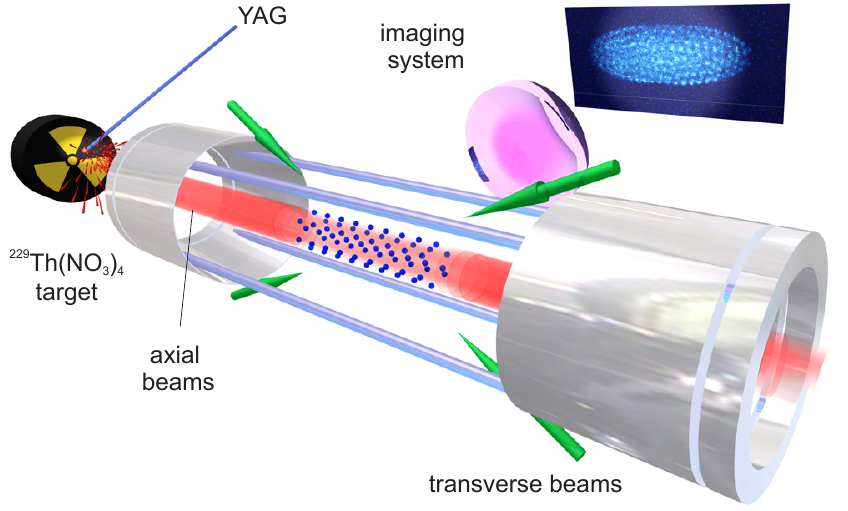}
\caption{Essential elements of the experimental setup.  See text for discussion.}
\label{fig:setup}
\end{figure}

In this Letter, we report an important step toward these goals: laser cooling, to crystallization, of $^{229}$Th$^{3+}$ in a linear Paul trap, with the ultracold ions stored for a large fraction of an hour.  Triply charged $^{229}$Th and $^{232}$Th ions are created via ablation [a 100 $\mu$J, 5 ns pulse at the third harmonic of a yttrium aluminum garnet (YAG) laser] of a thorium nitrate source and injected axially into a four-rod linear Paul trap, Fig. \ref{fig:setup}.  The design of the trap is motivated by the plasma nature of the Th$^{3+}$ ablation plume and is tailored for enhanced ion loading efficiency \cite{Campbell_thesis}.  The trap used here is of length 188 mm and radius 3.3 mm, with a 10 mm distance between the thorium nitrate source and the trap.

The trap is driven in the balanced mode at 4.3 MHz and single-ion radial trap frequencies for both isotopes are varied from 10 kHz - 1 MHz and the axial frequencies between 15 - 55 kHz during experimental sequences.  To create sufficient axial confinement with convenient dc voltages, two tube electrodes, of length 68 mm and 83 mm, sit coaxially with the rf potential and are separated by 22 mm at trap center.  The electrodes are biased at 3-50 V.  A tapered geometry of the tube inner diameters creates a potential gradient along the trap axis, inducing ion accumulation between the tubes.  For ion loading, a 6 mm long ring electrode biased at 250 V sits at each extreme end of the trap.  The ring electrode voltages are pulsed down to ground and back in sync with ion ablation, allowing Th$^{3+}$ entrance to the trap while blocking much of the slower moving Th$^{2+}$ and Th$^{+}$.  The mass-dependent stability of rf trapping is also utilized to avoid trapping pollutant ions and to remove pollutants from the trapped sample \cite{Campbell_thesis}.

Once trapped, laser cooling is sufficient to bring the ions from tens of thousands of Kelvin to crystallization, though overall yield is consistently $\lesssim$ 100 ions.  To increase this yield by 1 - 2 orders of magnitude, helium buffer gas of impurity $<$ 100 ppb is introduced into the vacuum system at a pressure of 10$^{-5}$ Torr, providing initial cooling to about room temperature.  The buffer gas is then removed, allowing for laser cooling to take place. The vacuum environment returns to a base pressure of $<10^{-11}$ Torr within a few minutes thereafter.

Laser cooling of trapped Th$^{3+}$ isotopes proceeds in two stages. First, transitions power broadened to several hundred times their natural linewidths are used for cooling ions from tens of thousands to tens of Kelvin.  The rather long excited state lifetimes of the $6D_{3/2}$ and $6D_{5/2}$  levels - of order 1 $\mu$s \cite{Safronova06}  -  are compensated for by the high laser intensities to allow cooling to overcome the heating associated with micromotion.  Subsequently, laser intensities which power broaden transitions to only several natural linewidths are utilized in cooling the ions into the mK regime.

\begin{figure}[t]
\includegraphics[scale=.98]{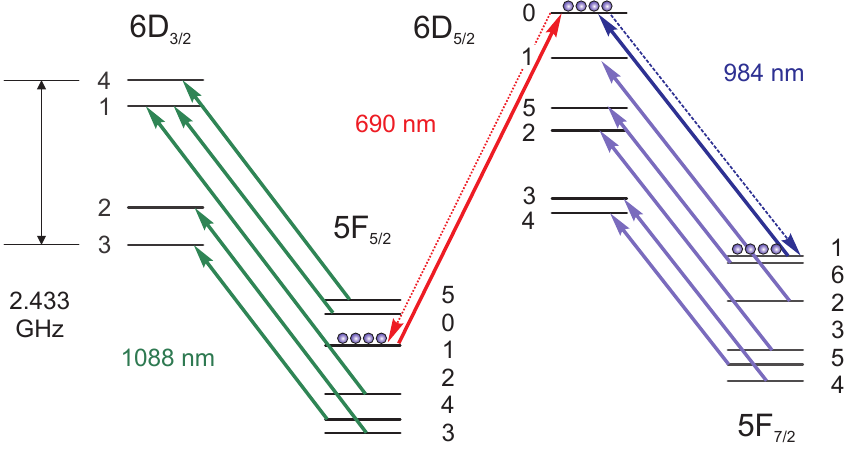}
\caption{Various optical transitions between hyperfine levels in $^{229}$Th$^{3+}$ employed during laser cooling.  See text for details.}
\label{fig:hyperfine}
\end{figure}

To scatter light from and laser cool $^{229}$Th$^{3+}$ in the steady state, multiple optical fields are required to repump from the various ground ($5F_{5/2}$) and metastable ($5F_{7/2}$)  hyperfine levels.  Fig. \ref{fig:hyperfine} illustrates relevant energy levels of $ ^{229}$Th$^{3+}$ and the scheme used in this work; the 1088 nm transitions are used in the high-temperature regime and the closed $|5F_{5/2}, F=1\rangle$, $|6D_{5/2}, F= 0\rangle$, $|5F_{7/2}, F=1\rangle$ $\Lambda$ system is used in the  low-temperature regime.  In this case, the requirement of multiple finely tuned optical frequencies is relaxed, as only repumping by 1088 nm and 984 nm fields is needed to correct for off-resonant excitation into the $6D_{5/2}$ manifold.  The $^{232}$Th$^{3+}$ system has no hyperfine structure, and therefore, no such complications arise \cite{Campbell09}.

The 1088 nm transitions of both isotopes are utilized for cooling the sample down to tens of Kelvin using two diode lasers.  The $^{229}$Th$^{3+}$ light is initially passed through a fiber-based electro-optic phase modulator (EOM), acquiring rf sidebands in the 0.1 - 1 GHz range.  In total, four 2 mm diameter beams, oriented in a common plane, converge on the ions at 45$^{\circ}$ angles to the trap axis.  The $^{232}$Th laser and the $^{229}$Th laser carrier and rf sidebands are tuned such that spectral components are located 20-80 MHz below relevant transition frequencies.  Each spectral component addressing $^{229}$Th$^{3+}$ carries about 1 mW of power per beam and the $^{232}$Th$^{3+}$ field carries about 20 mW per beam.  The lower $^{229}$Th power is a result of EOM optical insertion loss and the distribution of laser power over multiple rf sidebands.

For low-temperature cooling, transitions from the $5F_{7/2}$ manifolds to the $6D_{5/2}$ manifolds are excited with light from a frequency-stabilized 984 nm diode laser which passes through an EOM, acquiring rf sidebands in the frequency range 3-7 GHz.  The spectral components are placed 3 MHz below the resonant frequencies of the $^{232}$Th$^{3+}$ and relevant $^{229}$Th$^{3+}$ transitions.  The light is then directed to the ions with two 1 mm diameter beams counterpropagating along the trap axis.  Each beam contains about 20 $\mu$W of power per spectral component.  This beam orientation minimizes observed micromotion-induced Doppler broadening.  A nearly identical system at 690 nm is used to excite from the $5F_{5/2}$ manifolds to the $6D_{5/2}$ manifolds.  Each 690 nm beam carries about 60 $\mu$W of power in each relevant spectral component.  When ion temperature and Doppler width are reduced via initial 1088 nm laser cooling, the $\Lambda$ system fields become relevant, cooling both isotopes to crystallization.
\begin{figure}[b]
\includegraphics[scale=.98,angle=90]{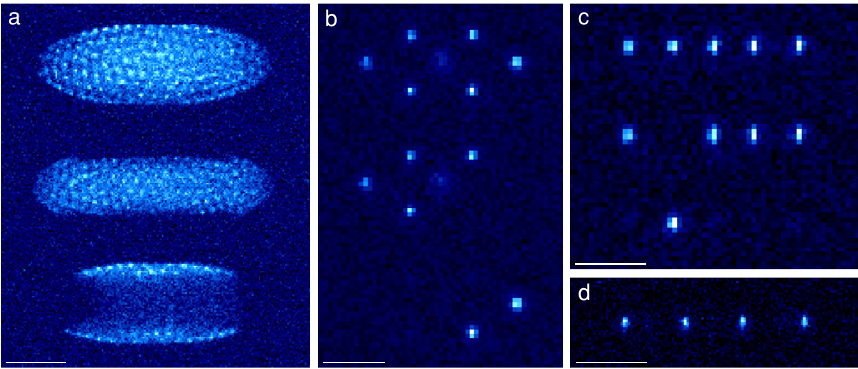}
\caption{Trapped $^{229}$Th$^{3+}$ and $^{232}$Th$^{3+}$ ions, laser cooled to crystallization.  (a)  Three successive images were taken of about 200 crystallized Th$^{3+}$ ions.  The top image was taken with both $^{229}$Th and $^{232}$Th 690 nm fields present.  The $^{232}$Th 690 nm field was then removed for the middle image.  The $^{232}$Th 690 nm field was then reintroduced and the $^{229}$Th 690 nm field simultaneously removed for the bottom image.  The larger mass-to-charge ratio of $^{232}$Th$^{3+}$ causes its radial accumulation in the outer shells of the crystal.  The short focal depth of the imaging system is apparent as most of the $^{232}$Th$^{3+}$ shell is out of focus.  The integration time is 1 s for all three images. (b), (c)  The same imaging protocol used in (a) is used on smaller samples.  The integration time is 2 s for all images.  (d)  A linear chain of four $^{229}$Th$^{3+}$ ions is shown.  The integration time is 3 s.  The scale bar in (a) is 500 $\mu$m and in (b)-(d) are 100 $\mu$m.}
\label{fig:crystals}
\end{figure}

During the laser cooling process, 984 nm fluorescence is used for ion detection.  Light is collected with a high numeric aperture (NA = 0.55) aspherical lens and directed onto a low-noise CCD camera.  The collection efficiency here is about 7\%.  Fig. \ref{fig:crystals} shows images of various Wigner crystals of $^{229}$Th$^{3+}$ and $^{232}$Th$^{3+}$.  Images were taken with different 690 nm fields present, controlling the level of 984 nm fluorescence from each of the two isotopes. The 10.5 GHz isotope shift is much larger than the broadened cooling transition widths of a few MHz, allowing for unambiguous isotopic identification.  The relative abundance of the two isotopes ($76\%$ $^{229}$Th and $24\%$ $^{232}$Th) in the thorium nitrate source is reflected in the observed composition of the dual-isotope crystals.  When low-temperature laser cooling is not present for one of the two isotopes, an ultracold phase is maintained due to the apparent sympathetic cooling.  For single ion excitation and detection, the most robust geometric configuration is the linear ion chain, such as those shown in Fig. \ref{fig:crystals}c,d.  Ions in this configuration are observed to survive in the trap for a large fraction of an hour.  The lifetime is likely limited by charge exchange collisions and chemical reactions with background molecules, processes enhanced by the large electron affinity of Th$^{3+}$ \cite{Churchill11}.

To determine and manipulate the internal states of ultracold $^{229}$Th$^{3+}$, optical transition frequencies between the multitude of hyperfine levels must be known at the level of a few MHz.  These frequencies are measured using optical pumping and fluorescence techniques via suitably engineered spectra at all three wavelengths.  To clearly resolve all optical transitions, the ions are sympathetically cooled by $^{232}$Th$^{3+}$ to tens of mK, narrowing the transition Doppler widths to a few MHz.  A single hyperfine level in the $5F_{5/2}$ or $5F_{7/2}$ manifold is left dark so that the ions are optically pumped into it. The 984 nm fluorescence is monitored on a camera while a probe field frequency is scanned. When this field is resonant with the targeted transition, the dark level is depopulated and a concomitant fluorescent signal is observed.  At maximal fluorescence, the value of the probe frequency is recorded with respect to the corresponding $^{232}$Th$^{3+}$ transition.

The magnetic dipole ($A$) and electric quadrupole ($B$) hyperfine coefficients of all four levels and the relative isotope shifts of all three transitions are extracted directly from the recorded data via a global least-squares fit containing expressions of the form \cite{Schwartz55}
\small
\begin{align*}
 \delta E_{e,g}& = \frac{K_e}{2} A_e + \frac{\frac{3}{2} K_e (K_e+1) - 2 I (I+1) J_e (J_e + 1)}{4 I (2 I - 1) J_e (2 J_e-1)} B_e \\
&- \frac{K_g}{2} A_g - \frac{\frac{3}{2} K_g (K_g+1) - 2 I (I+1) J_g (J_g + 1)}{4 I (2 I - 1) J_g (2 J_g-1)} B_g + \Delta \, ,
\end{align*}
\normalsize
where the subscript $g(e)$ indicates the ground (excited) state, $K_i=F_i(F_i+1)-I(I+1)-J_i(J_i+1)$, and $\Delta$ represents the relative isotope shift common to all optical transitions between a given pair of hyperfine manifolds.  Results of the fit are presented in Table \ref{tab:table1}.

Overall, about $20\%$ of the uncertainty is statistical for each extracted value, while the remaining $80\%$ is due to uncertainties in Zeeman splittings combined with optical pumping effects, laser cooling dynamics, and light shifts.  By taking the unweighted average ratio of the four measured $B$'s to the corresponding calculated electronic hyperfine matrix elements of Ref. \cite{Berengut09}, a value of the spectroscopic nuclear electric quadrupole moment $Q$ = 3.11(16) eb is obtained.  Here, the uncertainty is determined by the few percent accuracy of the atomic structure calculations.  No accurate calculations are currently available to extract the value of the nuclear magnetic dipole moment, although they are possible \cite{Safronova99}.  Our value for $Q$ may be compared with the previous value of $Q$ = 4.3 eb $\pm$ 20\% measured in Th$^+$ \cite{Gerstenkorn74}.

The most recent and precise published measurement of the $^{229}$Th isomer energy is 7.6(5) eV \cite{Beck07}.  In order to span $\pm\,3\,\sigma$ in the search for this nuclear level, direct optical excitation of the isomer in trapped cold ions may not be a viable method, given available UV sources and the large energy uncertainty.  Instead, the electron-bridge (EB) process may be utilized \cite{Campbell09}.  In this case, hyperfine-induced mixing of the ground and the isomer nuclear manifolds opens electric-dipole transitions between the two.  The mixing is expected to be the strongest for the $S$-electronic states, as the electron probability density at the nucleus is highest.  For example, considering only the $7S_{1/2}$ and $8S_{1/2}$ electronic orbitals in first-order perturbation theory,
\small
\begin{equation*}
 \overline{|7S_{1/2}, m\rangle} \approx |7S_{1/2}, m\rangle + \frac{\langle 8S_{1/2}, g|H_{int}|7S_{1/2}, m\rangle}{ E_{7S, m} - E_{8S, g}} |8S_{1/2}, g\rangle
\end{equation*}
\normalsize
where $H_{int}$ is the electron-nucleus interaction Hamiltonian and $g(m)$ indicates the nuclear ground (isomer) level.  The $|8S_{1/2},g\rangle$ admixture, with expected amplitude of order $10^{-5}$ \cite{Porsev10}, couples to the $|7P_{1/2},g\rangle$ level via electric-dipole radiation of frequency $(E_{7S,m} - E_{7P_{1/2},g})/\hbar$ (see Fig. \ref{fig:grotrian}).  This shifts the spectral interrogation region from the challenging 130-200 nm range to the manageable 250 - 800 nm range where high-power coherent light sources are available. It should be noted that this electron-bridge transition is predicted to be substantially stronger than the nuclear $M1$ transition \cite{Porsev10}.
\begin{table}[t]\small
\caption{\label{tab:table1} Measured $^{229}$Th$^{3+}$ hyperfine constants $A$ and $B$ and relative isotope shifts from $^{232}$Th$^{3+}$.  All units are MHz and all uncertainties are one sigma.}
\begin{ruledtabular}
\begin{tabular}{lccccc}
\textrm{Valence}&
\textrm{}&
\textrm{}&
\textrm{}&&
\textrm{Isotope}\\
\textrm{Orbital}&
\textrm{$A$}&
\textrm{$B$}&&
\textrm{Transition}&
\textrm{Shift}\\
\colrule
$5F_{5/2}$ & 82.2(6) & 2269(6) & \quad \quad &$5F_{5/2} \leftrightarrow 6D_{3/2}$ & -9856(10)\\
$5F_{7/2}$ & 31.4(7) & 2550(12) & &$5F_{5/2} \leftrightarrow 6D_{5/2}$ & -10509(7)\\
$6D_{3/2}$ & 155.3(12) & 2265(9) & &$5F_{7/2} \leftrightarrow 6D_{5/2}$ & -9890(9)\\
$6D_{5/2}$ & -12.6(7) & 2694(7) \\
\end{tabular}
\end{ruledtabular}
\end{table}

In general, the optical search time required to find the isomer transition energy can be estimated as $T \sim \Delta \omega / \Omega^2$, where $\Omega$ is the resonant optical excitation Rabi frequency and $\hbar \, \Delta\omega$ is the energy range over which the search is conducted.  Considering the $|7P_{1/2},g\rangle \leftrightarrow |7S_{1/2}, m\rangle$ electron-bridge, an effective electric-dipole moment, $d_{EB}$, is predicted to be $\approx 2\times10^{-5} \, e a_0$  \cite{Porsev10}.  For an ion in the $|7P_{1/2}, g\rangle$ level, placed at the center of a Gaussian light beam of power $P$ and waist $r_0$,  $\Omega^2 = (4 \,  P \,{d_{EB}}^2)/(\pi \, \epsilon_0 c \, {r_0}^2 \hbar^2)$ so that $T \sim (\pi \, \epsilon_0 c \, {r_0}^2 \hbar^2 \Delta\omega) / (4 \,  P \,{d_{EB}}^2)$.

The spectral range in which this transition is likely to exist (250 nm-800 nm) is completely accessible with a mode-locked Ti:Sapp oscillator at the fundamental (1 W), second (150 mW), and third (20 mW) harmonics, and with an optical parametric oscillator in the visible spectrum (200 mW).  By focusing ultra-fast excitation pulses from these or similar light sources to a waist $r_0 \approx 5$ $\mu m$ and onto a single crystallized $^{229}$Th$^{3+}$ ion, we estimate that the isomer level can be excited from the $|7P_{1/2}, g\rangle$ level in tens of hours of ion illumination.  Alternatively, $|7S_{1/2},g\rangle \leftrightarrow |7S_{1/2}, m\rangle$ two-photon excitation through the $|7P_{1/2}, g\rangle$ level may be a suitable approach for the search.  It should be noted that the ionization potential of Th$^{3+}$ is 29 eV, making multi-photon ionization with these protocols unlikely.

Once the isomer level is found, a single ion within a linear crystallized chain (Fig. \ref{fig:crystals}c,d) can been excited to the isomer manifold and its hyperfine structure and isomeric level shifts may be accurately measured.  This knowledge would allow both for unambiguous identification of the isomer level and for an empirical determination of the isomer transition sensitivity to $\alpha$ variation \cite{Berengut09}.

In summary, we have produced laser-cooled crystals of triply charged $^{229}$Th in a linear Paul trap.  Our work opens an avenue toward location and precise measurement of the nuclear transition energy in a single trapped, cold ion.  Laser excitation of this nuclear transition may lead to new levels of precision in timekeeping and tests of temporal variation of fundamental constants.

We thank J. Z. Blumoff and D. E. Naylor for their laboratory contributions, L. R. Churchill, M. V. DePalatis, and M. S. Chapman for assistance at an early stage of this work, and D. N. Matsukevich, M. Safronova, J. Berengut, and S. Porsev for discussions. This work was supported by the Office of Naval Research and the National Science Foundation.

\emph{Note added.}---While writing this manuscript, we became aware of Ref. \cite{Bemis1988}, where the value of $Q$ = 3.149$\pm$0.032 eb for $^{229}$Th was determined via Coulomb excitation of the nucleus, in good agreement with our result.

\end{document}